\documentclass[%
 reprint,
 amsmath,amssymb,
 aps,
]{revtex4-2}

\usepackage[title]{appendix}
\usepackage{graphicx}
\usepackage{dcolumn}
\usepackage{bm}
\usepackage{xcolor} 

\begin{document}

\preprint{APS/123-QED}

\title{Symmetry Broken Vectorial Kerr Frequency Combs from Fabry-Pérot Resonators}

\author{Lewis \surname{Hill$^{1,2}$}}
\email{lewis.hill@mpl.mpg.de}
\author{Eva-Maria \surname{Hirmer$^{1,3}$}}
\author{Graeme \surname{Campbell$^{2}$}}
\author{Toby \surname{Bi$^{1,3}$}}
\author{Alekhya \surname{Ghosh$^{1,3}$}}
\author{Pascal \surname{Del'Haye$^{1,3}$}}
\author{Gian-Luca \surname{Oppo$^{2}$}}
\affiliation{$^1$Max Planck Institute for the Science of Light, Staudtstr. 2, 91058 Erlangen, Germany\\$^2$SUPA \& CNQO, Department of Physics, University of Strathclyde, 107 Rottenrow, Glasgow, G4 0NG, UK\\$^3$Department of Physics, Friedrich Alexander University Erlangen-Nuremberg, 91058 Erlangen, Germany}

\begin{abstract}
Optical frequency combs find many applications in metrology, frequency standards, communications and photonic devices. We consider field polarization properties and describe a vector comb generation through the spontaneous symmetry breaking of temporal cavity solitons within coherently driven, passive, Fabry-Pérot cavities with Kerr nonlinearity. Global coupling effects due to the interactions of counter-propagating light restrict the maximum number of soliton pairs within the cavity - even down to a single soliton pair - and force long range polarization conformity in trains of vector solitons.
\end{abstract}

\maketitle
\begin{widetext}
\section*{\label{sec:Intro}Introduction}

Commonly described as ``Rulers for Light", optical frequency combs \cite{hansch2006nobel, del2007optical, pasquazi2018micro, fortier201920} and their generation are topics with wide-reaching research interest. This interest is especially owed to their diverse application in metrology, optical communications \cite{lundberg2020phase} and novel photonic devices \cite{pasquazi2018micro}. They are used for example in high precision spectroscopy, optical atomic clocks, and navigation systems. One particular method for producing an optical frequency comb utilizes dissipative Temporal Cavity Solitons (TCS) in optical micro-resonators.

We concern ourselves here not only with the generation of TCS but also with combining them with the phenomena known as Spontaneous Symmetry Breaking (SSB). Broadly defined, SSB describes the situation when two or more properties of a system suddenly change from displaying an equal property or state (symmetric) to having these states becoming unequal  (asymmetric) following an infinitely small change to some control parameter.

The SSB of light in Kerr resonators has seen a flurry of study and interest in the past half-decade, being predicted and observed for systems with counter-propagating light components \cite{kaplan1981enhancement, kaplan1982directionally,wright1985theory, del2017symmetry, woodley2018universal, hill2020effects, woodley2021self, cui2022control, bitha2023complex}, for orthogonally polarized light components \cite{moroney2022kerr, garbin2020asymmetric, xu2021spontaneous, xu2022breathing, quinn2023random, coen2023nonlinear, quinn2023towards}, and also very recently for a single system combining both counter-propagating light and two orthogonal polarization components together \cite{hill20224}, and a system containing multiple resonators \cite{alekhya}.

Although recent interest about TCS in Kerr resonators mainly focused on ring geometries, frequency combs based on TCS have also been found in Fabry–Pérot resonators, or linear cavities \cite{cole2018theory,wildi2022soliton,campbell2023dark}. Figure \ref{fig:Schematic} shows a basic schematic of this type of resonator, and this is the configuration of interest in this paper. One can see that it is comprised primarily of two highly reflective mirrors, which bounce light back and forth between them. Here we consider that the space between the mirrors is filled with a Kerr, or $\chi^{3}$, nonlinear medium.

\begin{figure}
    \centering
    \includegraphics[width=0.5\columnwidth]{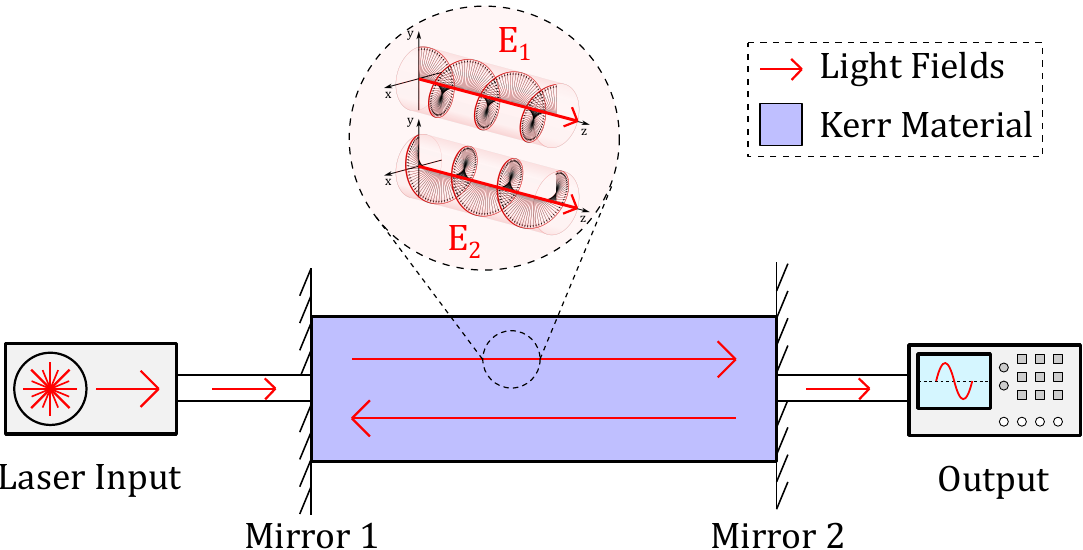}
    \caption{Schematic of a Kerr Fabry-Pérot cavity. The cavity is bounded by two highly reflective mirrors, with a Kerr nonlinear medium filling the space between them. Linearly polarized laser input is introduced via mirror 1, allowing the light to enter the cavity. Once inside, the light undergoes many reflections between the mirrors before eventually exiting. The intracavity field is analyzed by decomposing it into its left- and right-circularly polarized components.}
    \label{fig:Schematic}
\end{figure}

There has been great success in observing the SSB of a pair of vector TCS in Kerr \textit{ring} resonators \cite{xu2021spontaneous, xu2022breathing}, where the TCS have orthogonal polarizations, but similar phenomena in \textit{linear} Fabry–Pérot cavities, Fig. \ref{fig:Schematic}, has remained unreported. This is despite recent separate experimental observations in Fabry–Pérot cavities of both the SSB of flat solutions \cite{moroney2022kerr} and of scalar TCS \cite{wildi2022soliton}. Here we outline the SSB of a pair of vector TCS in Fabry–Pérot cavities.

\section*{Results and Discussions}

To model the intra-cavity, and slower, temporal dynamics of a field propagating in a Fabry-Pérot resonator with consideration for its polarisation, Fig. \ref{fig:Schematic}, we derive, see the Methods section, a system of two coupled, generalised, Lugiato-Lefever Equations (LLEs) \cite{lugiato1987spatial} with fast-time averaged terms. Here we rewrite the final integro-partial differential equations from the Methods section for the complex amplitudes $E_\pm$ of the circularly polarization components

\begin{equation}\label{Model}
    \frac{\partial E_\pm}{\partial t} = E_{in} - E_\pm - i\theta E_\pm - i \eta \frac{\partial^2E_\pm}{\partial\tau^2} + i\left(A|E_\pm|^2E_\pm + B|E_\mp|^2E_\pm + 2A\langle|E_\pm|^2 \rangle E_\pm + B\langle |E_\mp|^2\rangle E_\pm + B\langle E_\pm E_\mp^*\rangle E_\mp\right),
\end{equation}

\noindent where $E_{in}$ is the input pump, $\theta$ the cavity detuning, $\eta$ controls the type of dispersion (with its value $\pm1$ referring to normal or anomalous dispersion, respectively),  $t$ and $\tau$ are both temporal variables but on relative slow and fast time-scales, respectively \cite{haelterman1992dissipative}, and $A$ and $B$ control the strengths of self- and cross-phase modulation effects, \cite{hill2020effects, hill20224}. The terms with angled brackets, i.e. $\langle f \rangle$, represent the temporal averages of the encapsulated functions over a single resonator round-trip, and are defined by the following:

\begin{equation}\label{Averages}
    \langle f \rangle=\frac{1}{t_R}\int_{-t_R/2}^{\;t_R/2} f\left(\tau\right) \,d\tau.
\end{equation}

Comparing it with the equations of Cole et al. \cite{cole2018theory, campbell2023dark}, we see that the effects of the polarisation components cause not only additional cross-phase modulation terms (and their averages) but also a further additional term, the final term of Eq. \eqref{Model}, which is caused by an energy exchange between the two circular components of each beam \cite{pitois2001nonlinear,hill20224}. We note that, if so desired, one may transition Eq. \eqref{Model} to the linear polarization basis through its manipulation around the substitution of $E_\pm=\frac{1}{\sqrt{2}}\left(E_x\pm iE_y\right)$. Note also that the main symmetry of Eqs. \eqref{Model} is the invariance under the exchange of the plus and minus indexes. We recognise the potential for further generalisation and study of this system when taking into account higher order dispersion effects \cite{blanco2016pure,li2020experimental, li2020experimental, anderson2022zero, zhang2022microresonator, bi2023pure}, which could result in additional localised structures to those discussed here.

\subsection{\label{sec:HSS}Homogeneous Stationary States}

The homogeneous (meaning here unchanging over the $\tau$ domain) and stationary (meaning here unchanging with $t$) solutions (HSS) to Eq. \eqref{Model} are obtained by setting $\partial E_\pm/\partial t$ and $\partial^2 E_\pm/\partial \tau^2$ both to zero. We may further make use of the fact that when a function $f(\tau)$ is homogeneous over $\tau$ its average over the $\tau$ domain is equal to its value at a single point - that is to say Eq. \eqref{Averages} becomes trivially $f$. Hence, under homogeneous and stationary conditions, and following suitable algebraic manipulation, Eq. \eqref{Model} becomes

\begin{equation}\label{HSS}
    |E_\pm|^2 = \frac{|E_{in}|^2}{1+\left(3A|E_\pm|^2 + 3B|E_\mp|^2 - \theta\right)^2}.
\end{equation}

\noindent Equation \eqref{HSS} is identical in its mathematical form, although not physical meaning, to the homogeneous stationary states of two linearly polarised fields counter-propagating a Kerr ring resonator \cite{kaplan1981enhancement, kaplan1982directionally, woodley2018universal, campbell2022}, or of two orthogonally polarized fields co-propagating a Kerr ring resonator \cite{geddes1994polarisation, hill2020effects, garbin2020asymmetric}. Due to its mathematical analogies to other such systems, Eq. \eqref{HSS} has been studied extensively, and we shall not repeat that analysis here.

One very important property of Eq. \eqref{HSS} worthy of re-noting here however is its ability to display spontaneous symmetry breaking (SSB). In this context, SSB describes the phenomena where the two circulating fields $E_\pm$ go from having symmetric intensities, $|E_+|^2=|E_-|^2$, to having asymmetric intensities, $|E_+|^2\neq|E_-|^2$, upon an infinitely small change to input conditions (input conditions such as the input intensity $|E_{in}|^2$ or the cavity detuning, $\theta$, for example). With Eq. \eqref{HSS} and its SSB largely explored elsewhere, we display for the later benefit of the reader in Fig. \ref{fig:HSSSSB} (a) \& (b), examples of SSB in input intensity and cavity detuning scans for Eq. \eqref{HSS}, respectively, for self- and cross-phase modulation values $A=2/3$ and $B=4/3$ -- the values for silica glass fibers. These values give a very general $B/A$ ratio \cite{hill2020effects}.

\begin{figure}
    \centering
    \includegraphics[width=0.5\columnwidth]{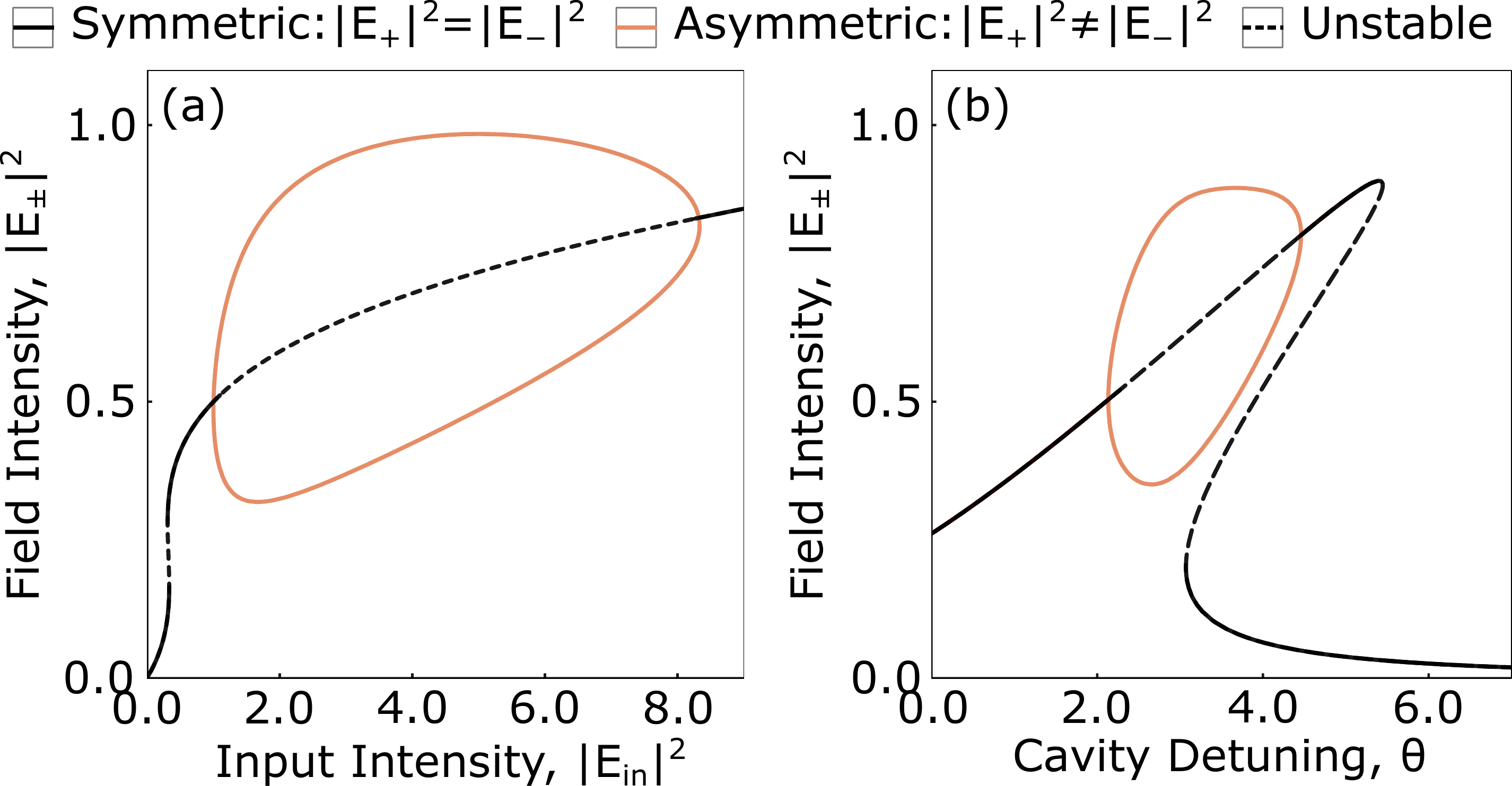}
    \caption{Spontaneous Symmetry Breaking of homogeneous solutions. Variations in the circulating fields' intensities as (a) input intensity, and (b) cavity detuning, are scanned. For this figure we use $A=2/3$ and $B=4/3$, with $\theta=2$ for (a) and $|E_{in}|^2=0.9$ for (b). Observe how in both panels ``red'', asymmetric, solutions emerge spontaneously from ``black'', symmetric, solutions at points where SSB occurr via pitchfork bifurcations. Note regions where the symmetric solution line is unstable to perturbations in $t$ (dashed line) on the middle branches of optical bistability, and also between SSB points.}
    \label{fig:HSSSSB}
\end{figure}

In the next section we proceed to describe the stability of the homogeneous and stationary states and hence their susceptibility to oscillations following fast- and/or slow-time perturbations.

\subsection{\label{sec:LinStab}Linear stability analysis of homogeneous stationary states}

To assess the system's susceptibility to temporal perturbations, both on $t$ and $\tau$, we performed a linear stability analysis on the modal expansion of Eq. \eqref{ModalApp}. General mathematical methods to find instability thresholds in single polarization Fabry-Pérot resonators have recently been established \cite{firth2021}. Here we linearised the modal equations around a homogeneous stationary solution $E_{\pm HS}$ with $U_n=E_{\pm HS}\delta_{n,0}+\delta U_n$, which, without loss of generality, had its phases adjusted such that it is real. We find that the four linear stability eigenvalues have the form

\begin{equation}\label{Eigenvalues}
    \lambda=-1\pm\frac{\sqrt{-\alpha_+\beta_+-\alpha_-\beta_--\frac{1}{6}\left(1-\delta_{n,0}\right)C\pm S}}{\sqrt{2}}
\end{equation}

\noindent with

\begin{equation}
    S=\sqrt{(\alpha_+\beta_+-\alpha_-\beta_-)^2+\alpha_+\alpha_-C+\left(1-\delta_{n,0}\right)\frac{C}{3}\left(\alpha_+\beta_++\frac{1}{3}\beta_+\beta_-+\alpha_-\beta_-\right)},
\end{equation}

\noindent where

\begin{equation}
    \begin{split}
        \alpha_\pm=&-\theta+\eta_n+3AE_{\pm HS}^2+B\left(2+\delta_{n,0}\right)E_{\mp HS}^2,\\
        \beta_\pm=&-\theta+\eta_n+A\left(5+4\delta_{n,0}\right)E_{\pm HS}^2+B\left(2+\delta_{n,0}\right)E_{\mp HS}^2,\\
        C=&36B^2\left(1+3\delta_{n,0}\right)E_{+HS}^2E_{-HS}^2,
    \end{split}
\end{equation}

\noindent where

\begin{equation}
    \eta_n=\eta\left(n\frac{2\pi}{t_R}\right)^2=\eta k^2.
\end{equation}

In figure \ref{fig:EigenPlot}, we display the HSS, Eq. \eqref{HSS}, and their corresponding linear stability eigenvalues, Eqs. \eqref{Eigenvalues} from the Methods section, over a cavity detuning scan of Eq. \eqref{HSS} for anomalous dispersion, $k=2$, $|E_{in}|^2=1.3$, and $A=2/3$ and $B=4/3$.

Regarding solution stability, if the eigenvalues of Eq. \eqref{Eigenvalues} all have real parts less than zero, i.e. $Re(\lambda)<0$, then the solution in question is stable, whereas if at least one eigenvalue has a real part greater than zero, then the solution is unstable. If the stability eigenvalues with real parts greater than zero and imaginary part different from zero occur when $n=0$, then the solution will begin to oscillate over slow-time, if they occur when $n\neq0$ the solution will oscillate over fast-time (Turing patterns).

We discuss first the eigenvalues associated with the symmetric solution line of this plot. They confirm the instability of the middle branch of the tilted symmetric Lorentzian curve, and also that the upper branch of the same curve is unstable between the SSB bifurcations points as expected from previous work. Note, however, that the instability of the symmetric line extends even beyond the higher detuning value corresponding to the reverse SSB bifurcation owed to the unstable $n\neq0$ eigenvalues, resulting in symmetric Turing patterns. Similarly, the instability of the asymmetric solution line also extends beyond slow-time instabilities, again owed to the unstable $n\neq0$ eigenvalues resulting in asymmetric Turing patterns, since they remain unstable for detuning values larger than those where the $n=0$ eigenvalues have stabilised.

\begin{figure}
    \centering
    \includegraphics[width=0.5\columnwidth]{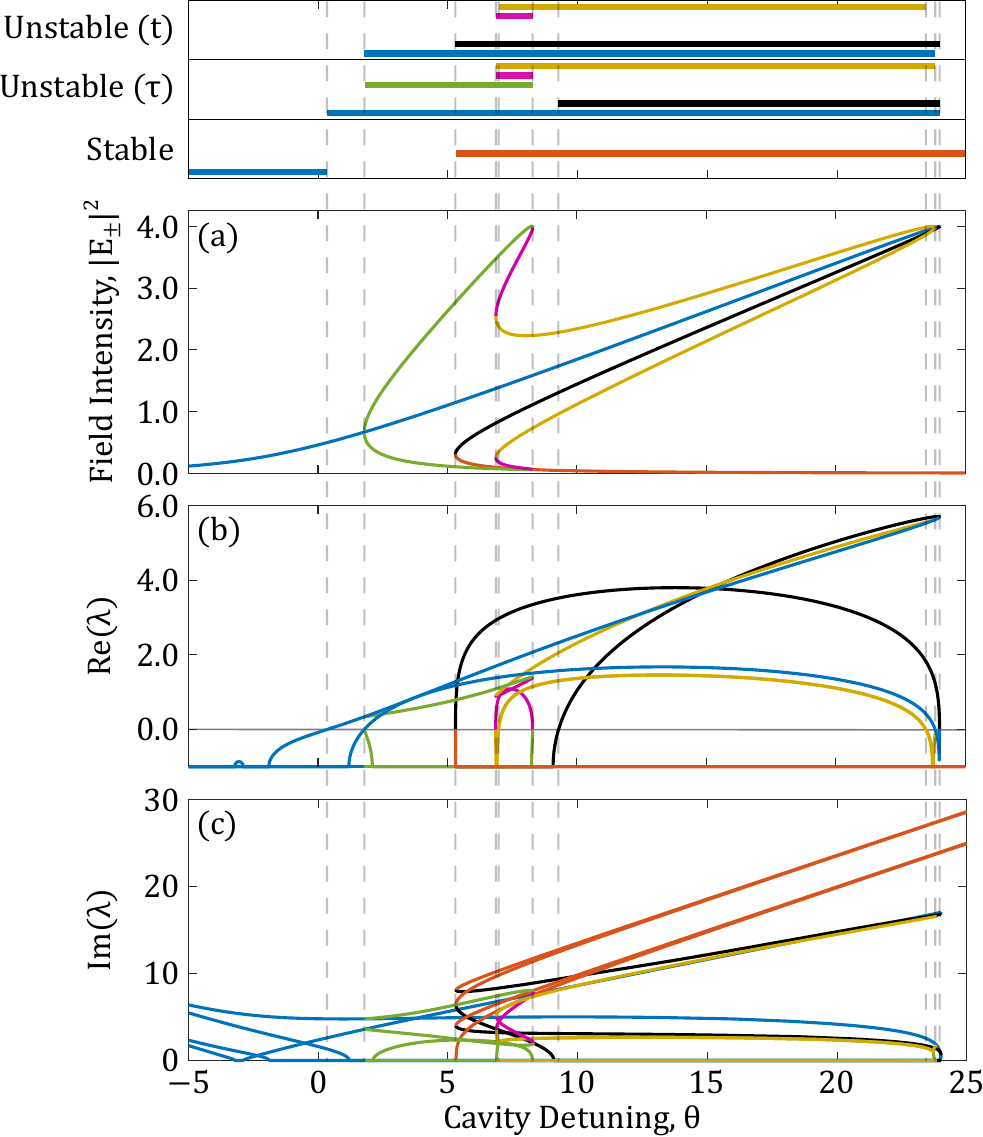}
    \caption{Homogeneous Stationary State Stability Analysis. Cavity detuning scan of (a) the homogeneous and stationary circulating field intensities (see Eq. \eqref{HSS}), (b) the Real part of the associated stability eigenvalues see Eq. \eqref{Eigenvalues}), and (c) the Imaginary part of the associated stability eigenvalues (see Eq. \eqref{Eigenvalues}), for anomalous dispersion, $k=1.9$, $|E_{in}|^2=4$, and $A=2/3$ and $B=4/3$. Eigenvalues in panels (b) and (c) use the colour corresponding to the different branches of HSS presented in panel (a) When at least one real part of the stability eigenvalues associated with a solution is greater than zero, $\textrm{Re}(\lambda)>0$, then the solution is unstable to perturbations. If the imaginary part of said eigenvalue is not equal to zero, $\textrm{Im}(\lambda)\neq0$, then the solution is susceptible to oscillations following perturbations, with the oscillations occurring over slow-time, $t$, fast-time, $\tau$, (Turing patterns), or both. The top panel summarises the stability of the various coloured HSS solutions of panel (a) based upon the stability eigenvalues of panels (b) \& (c).
    }
    \label{fig:EigenPlot}
\end{figure}

\begin{figure}
    \centering
    \includegraphics[width=0.9\columnwidth]{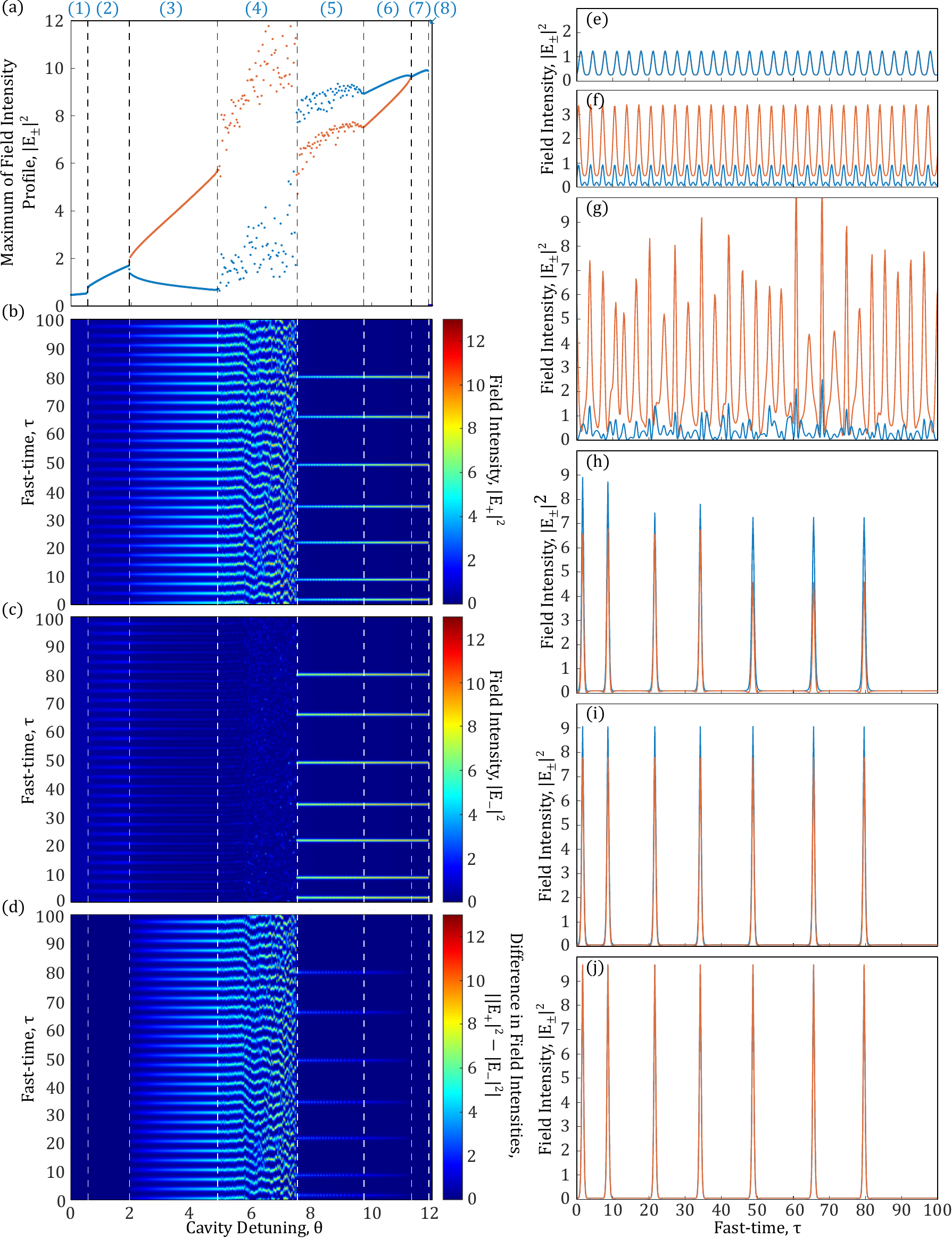}
    \caption{Vector soliton pair generation. Cavity detuning is scanned for an input intensity of $|E_{in}|^2=4$. Panel (a) shows the maximum value achieved across the cavity by the intensity profiles of the two fields, $|E_\pm|^2$ in red and blue, respectively. (b) and (c), respectively, show the full field intensity profiles, $|E_\pm|^2$, over the cavity, while (d) shows the difference between them. We note eight distinct regions in the panels (a) - (d), which we label (1) - (8). These regions are discussed in the main text. In panels (e) - (j) we show characteristic intensity profiles reflecting those existing in regions (2) - (7), respectively, with regions (1) and (8) being simply the homogeneous stationary profile.}
    \label{fig:RanScan}
\end{figure}

\subsection{\label{sec:Inhom}Inhomogeneous solutions: Patterns, Temporal Cavity Solitons}

Due to Eq. \eqref{Model}'s susceptibility to instabilities on the fast-time, we further explored the inhomogeneous solutions of the system. In this section, we always use $A=2/3$ and $B=4/3$, and here analyse situations only for anomalous dispersion, $\eta=-1$, -- situations related to normal dispersion, $\eta=+1$, will be discussed elsewhere. Figure \ref{fig:RanScan} shows in panels (a) -- (d) the variation in the field intensity profiles across the cavity as the cavity detuning is scanned, for a set input pump of $|E_{in}|^2 = 4$. We observe eight distinct regions of characteristic behaviours labeled in the figure as regions (1) -- (8). The detuning scans begin with the field intensity profiles following the symmetric HSS line (region 1), before forming symmetric Turing patterns across $\tau$ (region 2), in agreement with the linear stability analysis. A characteristic example of these symmetric patterns is presented in Fig. \ref{fig:RanScan} (e), its peaks grow in intensity as the cavity detuning is further increased. While these pattern states may be initially symmetric, this property later breaks spontaneously, resulting in a region of asymmetric patterns (region 3), with characteristics similar to that of Fig. \ref{fig:RanScan} (f). Progressing further still across the cavity detuning scan, the pattern's intensity profiles soon become unstable to chaotic oscillations (region 4), see the example in Fig. \ref{fig:RanScan} (g), before finally reaching a detuning value where the pattern states naturally form vector soliton-pair structures (region 5), see Fig. \ref{fig:RanScan} (h). One will note that initially the solitons in each of the field intensity profiles $|E_\pm|^2$ are both asymmetric and breathing (region 5). In Fig. \ref{fig:RanScan} (i), however, these asymmetric soliton pairs stop breathing and stabilise (region 6). The asymmetric soliton pairs eventually converge to symmetric profiles at the point starting region 7, as the example shows in Fig. \ref{fig:RanScan} (j). Finally, in region 8, the soliton pairs die, and the field intensity profiles return to the HSS line.

\subsection{\label{sec:Long}Long Range ``talking" of Soliton Pairs and Their Polarization Conformity}

In this section, we show that the angle bracketed terms of Eq. (\ref{Model}) have fascinating repercussions on the system's evolution. Although one of their effects has already been felt, the requirement to analyse separately the $n = 0$ stability eigenvalues, their wider effects have remained somewhat hidden until this section. We have stated that, mathematically, these angled terms amount to an average of the field intensity profiles across the cavity, but owed to this they cause a global coupling across the cavity \cite{campbell2022}. Related to the TCS pairs, these global coupling terms effectively mean that all the soliton pairs feel the impact of, and in turn influence, any others in the cavity.

One effect of this global communication, which was noted in Ref. \cite{cole2018theory} for the single polarization case, is that the maximum number of solitons existing within the cavity at any given time can be controllably limited by ones choice of system parameters. This ability could be extremely useful, for example, for applications that require robust temporal delays between subsequent soliton generations, such as for frequency combs. We show that this control on the maximum number of solitons present in the cavity is maintained for our vector soliton pairs.

\begin{figure}[b]
    \centering
    \includegraphics[width=0.5\columnwidth]{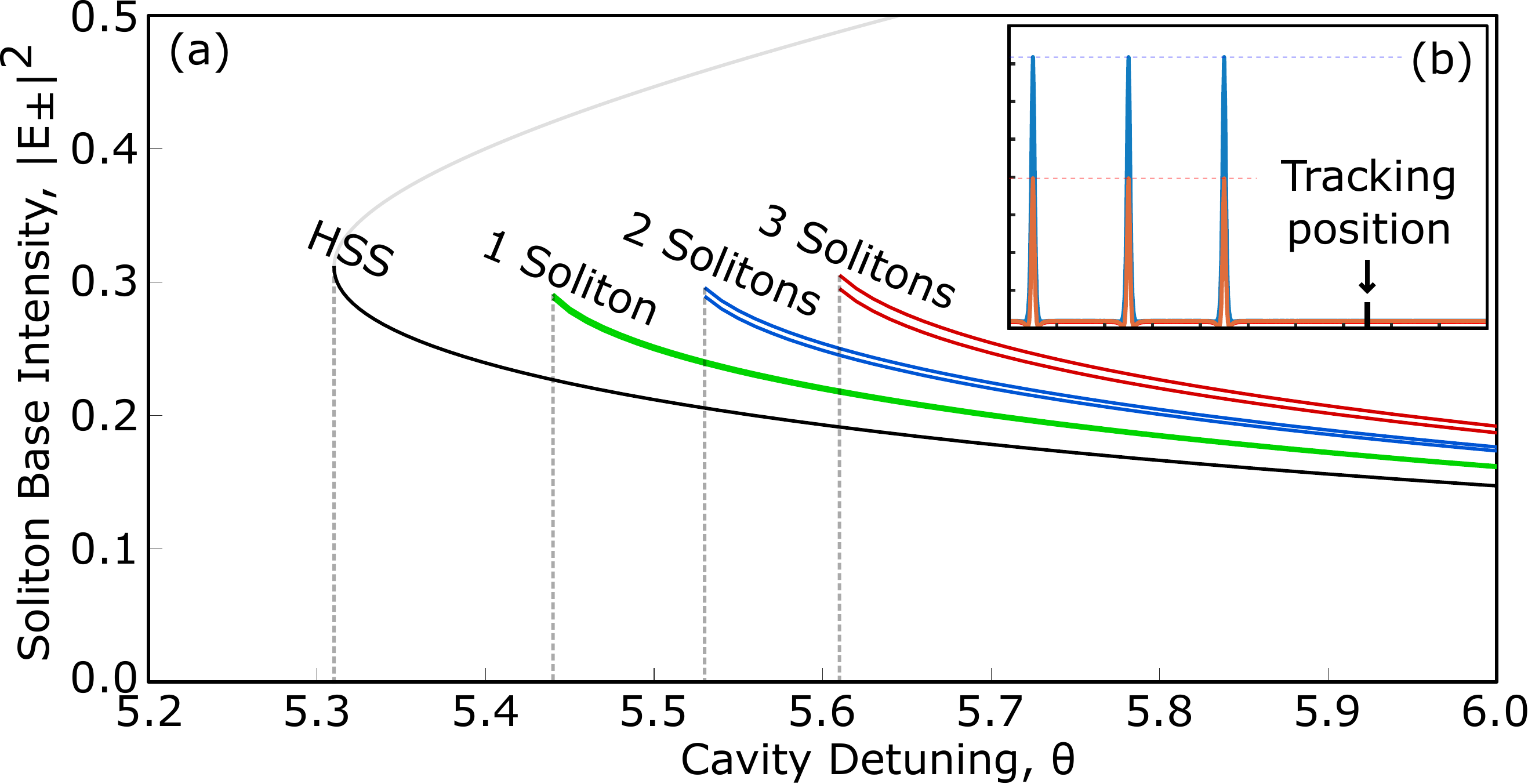}
    \caption{Existence of a viable soliton base. (a) Variation in the soliton-base intensity as the cavity detuning is varied, and also as the number of solitons within the cavity varies, for $|E_{in}|^2=4$. The solid black line displays the homogeneous stationary state line, the green line tracks the soliton-base when a single soliton exists within the cavity, the blue line tracks when two solitons are present, and the red line tracks when three solitons are present. Note that due to the asymmetry between the solitons existing in $E_\pm$ respectively there are two lines for each color tracking the soliton-base. In panel (b) we show a cavity containing three symmetry broken vector solitons and the position where we track the soliton-base for our scans in (a).}
    \label{fig:SolLife}
\end{figure}

Normally solitons in LLE systems, of the nature similar to those discussed in this manuscript, ``sit" upon the bottom branch of the symmetric HSS, and hence naturally lose viability when the HSS bottom branch ends. However, as displayed in Fig. \ref{fig:SolLife}, and owed to the $\langle f \rangle$ terms, these soliton bases are in fact elevated above the HSS in our system. One can also see that these $\langle f \rangle$ terms further have the effect of raising the cavity detuning value at which the solitons gain viability. This is because the $\langle f \rangle$ terms cause an effective detuning in the system \cite{cole2018theory, campbell2022}. Further still, one notes that the more solitons present in the cavity, the more by which this detuning limit is raised. This increase in the cavity dutuning value required to support additional solitons effectively limits the maximum soliton pair number supported at any one time, based upon the current cavity detuning value. In relation to Fig. \ref{fig:SolLife}, for example, at a cavity detuning of $\theta = 5.45$, the system can only support a maximum of a single pair of vector solitons, with the system being unable to support even a single additional pair until well after $\theta = 5.5$. This effect amounts to a system which can guarantee the generation of a maximum of a single vector soliton pair if so desired. Looking at the intensity profile of a single asymmetric soliton pair in the frequency domain, Fig. \ref{fig:SinSol}, we see the types of symmetry broken frequency combs that can be produced utilizing this phenomenon.

\begin{figure}[t]
    \centering
    \includegraphics[width=0.5\columnwidth]{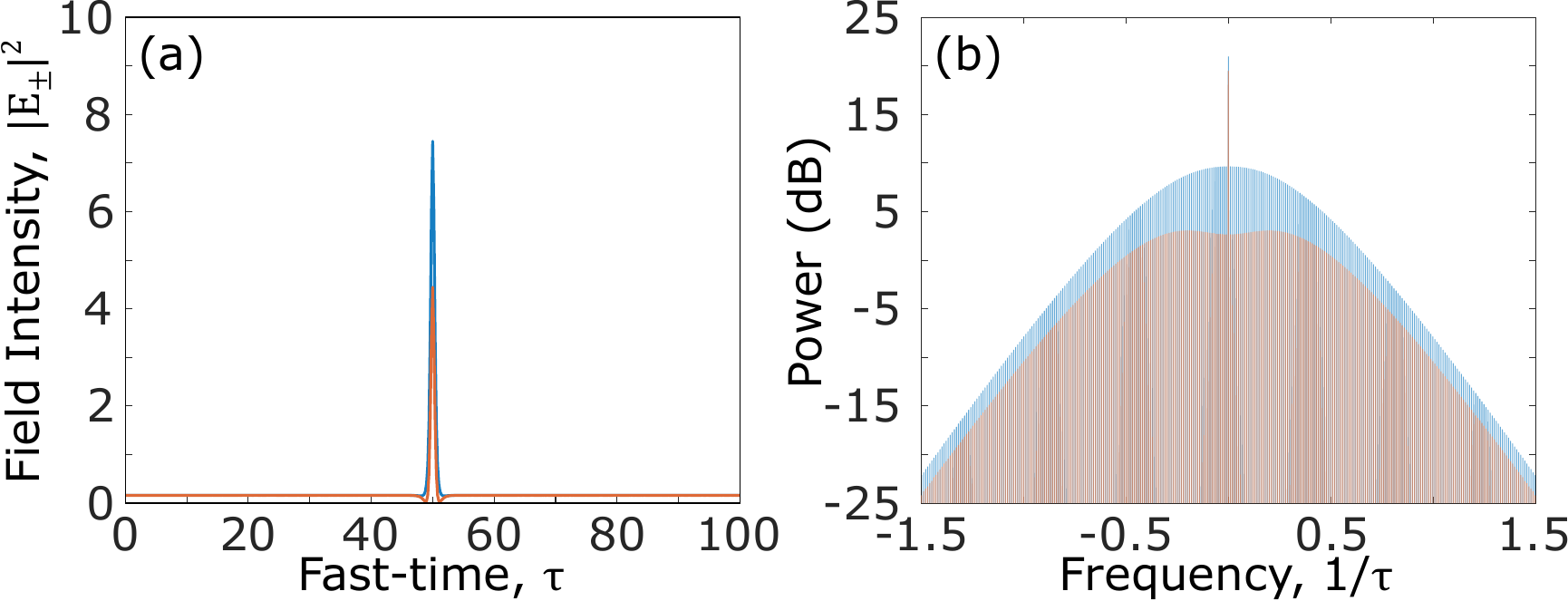}
    \caption{Examples of symmetry broken TCSs and dual frequency combs. Field intensity profiles across the cavity containing a single asymmetric soliton pair, panel (a), and the corresponding frequency combs in panel(b). Here $|E_{in}|^2=4$ and $\theta=6.2$.}
    \label{fig:SinSol}
\end{figure}

\begin{figure}[b]
    \centering
    \includegraphics[width=0.9\columnwidth]{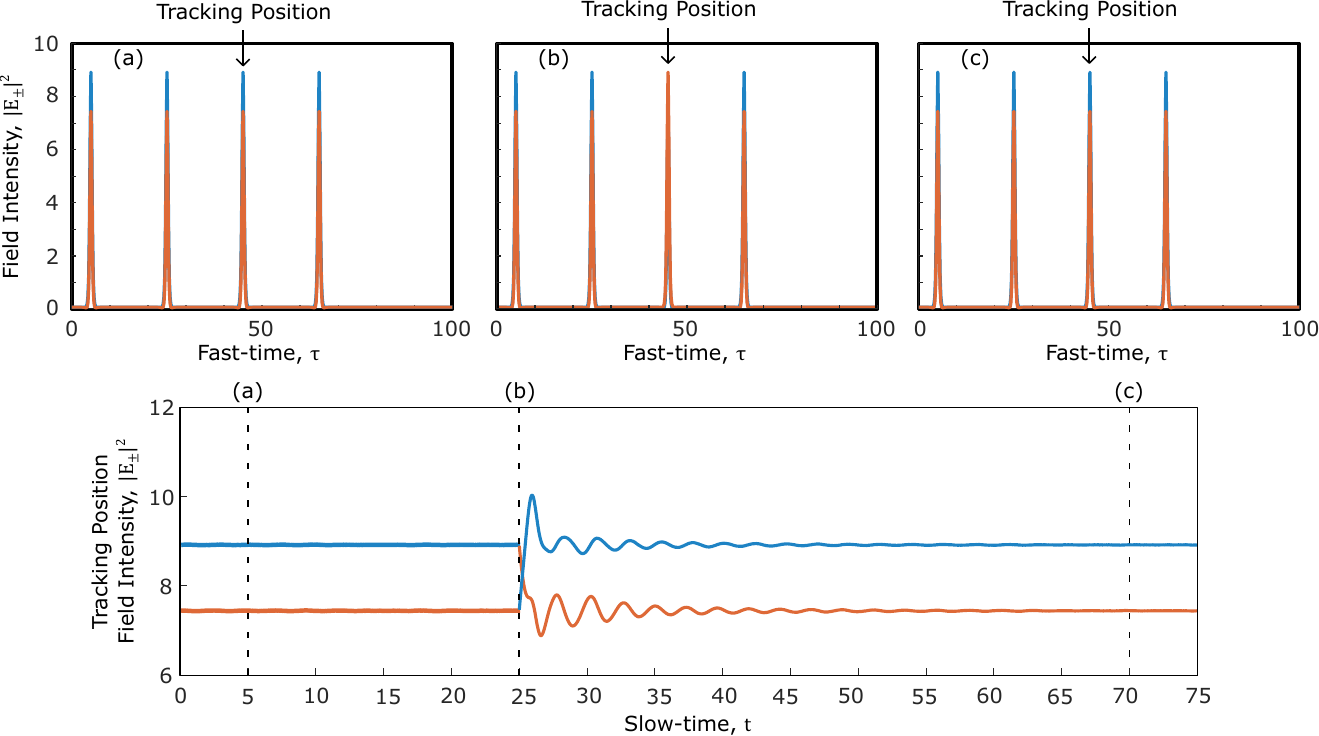}
    \caption{Long-range ``talking'' and conforming solitons. For $|E_{in}|^2=4$, $\theta=9$, the main panel of the figure tracks the paired soliton peaks at a set position within the cavity as they evolve over time. The stable initial cavity condition is formed by a train of four pairs of asymmetric solitons and is displayed in the smaller panel (a). At a slow-time value of $t=25$, the third soliton pair is forced to switch its polarization profiles, small panel (b). The system then evolves until it becomes stable once again, small panel (c). One should note that all four soliton pairs in (c) have conformed again to share the same dominant polarization.}
    \label{fig:SlowScan}
\end{figure}

The final effect of the angle-bracketed terms that will be discussed in this manuscript is perhaps the most interesting. The end result of this effect was already apparent in panels (h) and (i) of Fig. \ref{fig:RanScan}. One will note that the soliton pairs in these panels, both those breathing, and those that are stable, all share the same dominant polarisation although generated autonomously. This result is particularly intriguing when compared with that for soliton pairs produced in a Kerr ring \cite{xu2021spontaneous, xu2022breathing}, as opposed to a FP cavity, where there each asymmetric soliton pair's dominant polarization was independent of that held by any other pair. This meant that provided those soliton pairs were sufficiently spaced apart from each other, they could behave and be addressed independently. The fact that we always observe all simultaneous asymmetric soliton pairs in our cavities sharing the same dominant polarization is due to the global coupling terms that facilitate long-range interactions between all soliton pairs existing within the cavity, terms not present in the Kerr ring system, in particular the final term of Eq. \eqref{Model} caused by the energy exchange between the two circular components of each beam.

We find that the conformity of the soliton pairs to a globally dominant polarization is a strong and robust effect.  We demonstrate this in Fig. \ref{fig:SlowScan}. The main panel of the figure shows the slow-time evolution of a tracked point in the fast-time field intensity profiles, a point chosen to line up with the intensity maxima of one of the cavity soliton pairs. In panel (a) one sees the initial stable cavity condition made up of a train of four asymmetric soliton pairs, all sharing the same dominant polarization. This configuration is stable in the slow time. At time $t=25$ we attempt to force non-conformity on the system. We do this by splicing the field intensity profiles and swapping the field roles of one of the soliton pairs such that in the swapped pair (third) the opposing polarization is now dominant. Allowing the system to continue to evolve after this attempt to force non-conformity shows how the system evolves back to a state where soliton pair polarization conformity appears once again. This behaviour can have notable advantages in, for example, protecting communicated binary data encoded in the polarization of a chain of soliton pairs in noisy systems. The robustness of the polarization conformity is limited in the way one would likely expect - if the majority of soliton pairs have their polarization reversed, then the system shall evolve to the new dominant-on-average polarization until that polarization is conformed to across the system. If exactly half of the soliton pairs have their polarization inverted, then the eventual conformed polarization is determined by random perturbations.

\section*{Conclusions}

In conclusion, we have demonstrated the spontaneous symmetry breaking of a pair of vector temporal cavity solitons in Fabry–Pérot cavities. Our investigation not only revealed the stability and dynamical behavior of these vector solitons in various parameter regimes but also uncovered useful phenomena arising from the global coupling terms in our derived model. We showed that the maximum number of soliton pairs in the cavity can be self-limited with important implications for frequency comb applications. Furthermore, we discovered that asymmetric soliton pairs in our system exhibit a conformity in their dominant polarization states, highlighting the effects on soliton pair behaviour caused by the global coupling terms. These findings significantly expand our understanding of temporal cavity solitons and spontaneous symmetry breaking in Kerr cavities and may have extensions into studies around frequency combs from high order dispersion, thus openning up new avenues for further exploration and potential applications in optical metrology, pulse coding, optical communications and beyond.

\section*{\label{sec:Methods}Methods}

The nonlinear interaction terms in our model are based on those of Pitois et al. \cite{pitois2001nonlinear} where counter-propagation of polarization components is considered in an isotropic fibre. In this work we add dispersion and consider the reflective boundary conditions, the input pump, and the cavity detuning and other losses, all of which are inherent to a Fabry-Pérot resonator \cite{geddes1994polarisation, cole2018theory}. Then we generalise the equations for various self- ($A$) and cross-phase ($B$) modulation strengths, \cite{hill2020effects, hill20224}, resulting in the following Eqs. (\ref{R+}-\ref{L-}).

\vspace{0.3cm}
\noindent \textbf{Right Circular Polarisation:}\\
\indent\indent\indent Forward propagating, $u(t,\tau)$:
    \begin{equation}\label{R+}
        \frac{\partial u}{\partial t} + v_g \frac{\partial u}{\partial\tau}=E_{in} - u - i\theta u-i\eta\frac{\partial^2u}{\partial\tau^2}+i\left(A|u|^2u+B|v|^2u+2A|\bar{u}|^2u+B|\bar{v}|^2u+B\bar{u}\bar{v}^*v\right),
    \end{equation}
\indent\indent\indent Backward propagating, $\bar{u}(t,\tau)$:
    \begin{equation}\label{R-}
        \frac{\partial \bar{u}}{\partial t} - v_g \frac{\partial \bar{u}}{\partial\tau}=E_{in} - \bar{u} - i\theta \bar{u}-i\eta\frac{\partial^2\bar{u}}{\partial\tau^2}+i\left(A|\bar{u}|^2\bar{u}+B|\bar{v}|^2\bar{u}+2A|u|^2\bar{u}+B|v|^2\bar{u}+Buv^*\bar{v}\right),
    \end{equation}
\textbf{Left Circular Polarisation:}\\
\indent\indent\indent Forward propagating, $v(t,\tau)$:
    \begin{equation}\label{L+}
        \frac{\partial v}{\partial t} + v_g \frac{\partial v}{\partial\tau}=E_{in} - v - i\theta v-i\eta\frac{\partial^2v}{\partial\tau^2}+i\left(A|v|^2v+B|u|^2v+2A|\bar{v}|^2v+B|\bar{u}|^2v+B\bar{v}\bar{u}^*u\right),
    \end{equation}
\indent\indent\indent Backward propagating, $\bar{v}(t,\tau)$:
    \begin{equation}\label{L-}
        \frac{\partial \bar{v}}{\partial t} - v_g \frac{\partial \bar{v}}{\partial\tau}=E_{in} - \bar{v} - i\theta \bar{v}-i\eta\frac{\partial^2\bar{v}}{\partial\tau^2}+i\left(A|\bar{v}|^2\bar{v}+B|\bar{u}|^2\bar{v}+2A|v|^2\bar{v}+B|u|^2\bar{v}+Bvu^*\bar{u}\right),
    \end{equation}

\noindent where $u$ and $v$ are the forward propagating field components with right- and left-circular polarisations, respectively, with $\bar{u}$ and $\bar{v}$ being the backwards propagating variants (representing the reflections of fields $u$ and $v$ respectively), $E_{in}$ accounts for the input pump, $\theta$ is the cavity detuning (the difference between the frequency of the input pump laser and the closest cavity resonance frequency), $\eta$ controls the type of dispersion (with its value $\pm1$ referring to normal or anomalous dispersion, respectively), and $t$ and $\tau$ are both temporal variables but on relative slow and fast time-scales, respectively \cite{haelterman1992dissipative}. We also set the following boundary conditions

\begin{equation}\label{Boundary}
    \begin{split}
        u(t,0)&=\bar{u}(t,0),\;\;\;\;\;u(t,t_R/2)=\bar{u}(t,t_R/2),\\
        v(t,0)&=\bar{v}(t,0),\;\;\;\;\;v(t,t_R/2)=\bar{v}(t,t_R/2).
    \end{split}
\end{equation}

\noindent where $t_R$ is the round trip time of the resonator.

\noindent Drawing inspiration from Cole et al. \cite{cole2018theory}, we seek to combine Eq. \eqref{R+} \& \eqref{R-} and Eq. \eqref{L+} \& \eqref{L-}, respectively, together, such that Eqs.(\ref{R+}-\ref{L-}) reduce to a system with only two equations. To achieve this, we introduce the following modal expansions in terms of the modal amplitudes $\tilde{u}_n,\;\tilde{v}_n,\;\tilde{E}_{in}$:

\begin{equation}\label{ModalEx}
    \begin{split}
        &u=\sum_{n=-\infty}^{\infty} \tilde{u}_n(t)e^{in\frac{2\pi}{t_R}\tau},\;\;\;\bar{u}=\sum_{n=-\infty}^{\infty} \tilde{u}_n(t)e^{-in\frac{2\pi}{t_R}\tau},\\
        &v=\sum_{n=-\infty}^{\infty} \tilde{v}_n(t)e^{in\frac{2\pi}{t_R}\tau},\;\;\;\bar{v}=\sum_{n=-\infty}^{\infty} \tilde{v}_n(t)e^{-in\frac{2\pi}{t_R}\tau},\\
        &\indent\indent\indent\indent\indent\indent E_{in}=\sum_{n=-\infty}^{\infty} \tilde{E}_{in}\delta_{n,0}e^{in\frac{2\pi}{t_R}\tau},
    \end{split}
\end{equation}

\noindent where $\delta_{n,0}$ is the Kronecker delta function, and the modal amplitudes are given by

\begin{equation}\label{ModalExIn}
    \begin{split}
        \tilde{u}_n(t)&=\frac{1}{t_R}\int_{-t_R/2}^{\;t_R/2}\,d\tau\ u\left(t,\tau\right) e^{-in\frac{2\pi}{t_R}\tau}=\frac{1}{t_R}\int_{-t_R/2}^{\;t_R/2}\,d\tau\ \bar{u}\left(t,\tau\right) e^{in\frac{2\pi}{t_R}\tau},\\
        \tilde{v}_n(t)&=\frac{1}{t_R}\int_{-t_R/2}^{\;t_R/2}\,d\tau\ v\left(t,\tau\right) e^{-in\frac{2\pi}{t_R}\tau}=\frac{1}{t_R}\int_{-t_R/2}^{\;t_R/2}\,d\tau\ \bar{v}\left(t,\tau\right) e^{in\frac{2\pi}{t_R}\tau}.
    \end{split}
\end{equation}

\noindent These allow us to extend our field equations over a full round trip since

\begin{equation}\label{Boundary2}
    \begin{split}
        u(t,\tau)&=\bar{u}(t,-\tau),\;\;\;\;\;\bar{u}(t,\tau)=u(t,-\tau)\\
        v(t,\tau)&=\bar{v}(t,-\tau),\;\;\;\;\;\bar{v}(t,\tau)=v(t,-\tau).
    \end{split}
\end{equation}

Focusing momentarily alone on Eq. \eqref{R+}, we insert the modal expansions of Eqs. \eqref{ModalEx} \& \eqref{ModalExIn} to obtain
    \begin{equation}\label{long}
        \begin{split}
            \sum_{n=-\infty}^{\infty}\frac{\partial\tilde{u}_n}{\partial t}e^{in\frac{2\pi}{t_R}\tau}+ v_g \sum_{n=-\infty}^{\infty}in\frac{2\pi}{t_R}\tilde{u}_n e^{in\frac{2\pi}{t_R}\tau}&=\sum_{n=-\infty}^{\infty} \tilde{E}_{in}\delta_{n,0}e^{in\frac{2\pi}{t_R}\tau} - \sum_{n=-\infty}^{\infty} \tilde{u}_ne^{in\frac{2\pi}{t_R}\tau} \\
            &\;\;\;\;\;\;- i\theta \sum_{n=-\infty}^{\infty} \tilde{u}_ne^{in\frac{2\pi}{t_R}\tau}-i\eta\sum_{n=-\infty}^{\infty}\left(in\frac{2\pi}{t_R}\right)^2 \tilde{u}_ne^{in\frac{2\pi}{t_R}\tau}\\ &\;\;\;\;\;\;+i\sum_{n=-\infty}^{\infty}\sum_{n'=-\infty}^{\infty}\sum_{n''=-\infty}^{\infty}(A\tilde{u}_{n+n'-n''}\tilde{u}_{n'}^*\tilde{u}_{n''}+B\tilde{v}_{n+n'-n''}\tilde{v}_{n'}^*\tilde{u}_{n''}\\
            &\;\;\;\;\;\;+2A\tilde{u}_{-n+n'+n''}\tilde{u}_{n'}^*\tilde{u}_{n''}+B\tilde{v}_{-n+n'+n''}\tilde{v}_{n'}^*\tilde{u}_{n''}+B\tilde{u}_{-n+n'+n''}\tilde{v}_{n'}^*\tilde{v}_{n''})e^{in\frac{2\pi}{t_R}\tau},
        \end{split}
    \end{equation}

\noindent with a similar equation focusing on $\tilde{v}_n$. We can then decompose $\tilde{u}$ and $\tilde{v}$ into the product of two functions on distinct time scales by setting
\begin{equation}
    \tilde{u}_n(t)=U_n(t)e^{-iv_gn\frac{2\pi}{t_R}t},\;\;\;\; \tilde{v}_n(t)=V_n(t)e^{-iv_gn\frac{2\pi}{t_R}t}
\end{equation}

\noindent If we take an average of the decomposed Eq. \eqref{long} over an extended slow-time, $t$, interval then the exponential terms on the average vanish for $n''\neq n$. This turns Eq. \eqref{long} to

    \begin{equation}\label{Exp1}
        \begin{split}
            \sum_{n=-\infty}^{\infty}\frac{\partial U_n}{\partial t}e^{in\frac{2\pi}{t_R}\tau}&=\sum_{n=-\infty}^{\infty} \tilde{E}_{in}\delta_{n,0}e^{in\frac{2\pi}{t_R}\tau} - \sum_{n=-\infty}^{\infty} U_ne^{in\frac{2\pi}{t_R}\tau} \\
            &\;\;\;\;\;\;- i\theta \sum_{n=-\infty}^{\infty} U_ne^{in\frac{2\pi}{t_R}\tau}-i\eta\sum_{n=-\infty}^{\infty}\left(in\frac{2\pi}{t_R}\right)^2 U_ne^{in\frac{2\pi}{t_R}\tau}\\                 &\;\;\;\;\;\;+i\sum_{n=-\infty}^{\infty}\sum_{n'=-\infty}^{\infty}\sum_{n''=-\infty}^{\infty}(AU_{n+n'-n''}U_{n'}^*U_{n''}+BV_{n+n'-n''}V_{n'}^*U_{n''})e^{in\frac{2\pi}{t_R}\tau}\\
            &\;\;\;\;\;\;+i\sum_{n=-\infty}^{\infty}U_{n}e^{in\frac{2\pi}{t_R}\tau}\sum_{n'=-\infty}^{\infty}(2AU_{n'}U_{n'}^*+BV_{n'}V_{n'}^*)\\
            &\;\;\;\;\;\;+i\sum_{n=-\infty}^{\infty}V_{n}e^{in\frac{2\pi}{t_R}\tau}\sum_{n'=-\infty}^{\infty}(BU_{n'}V_{n'}^*),
        \end{split}
    \end{equation}

\noindent with again a similar equation focusing on $\tilde{v}_n$.

Finally, if we collapse the modal expansions by defining:

\begin{equation}
    E_+=\sum_{n=-\infty}^{\infty} U_ne^{in\frac{2\pi}{t_R}\tau}, \;\;\;\;\; E_-=\sum_{n=-\infty}^{\infty} V_ne^{in\frac{2\pi}{t_R}\tau},
\end{equation}

\noindent one obtains the used model, Eq. \eqref{Model}.
\begin{equation}\label{ModalApp}
    \frac{\partial E_\pm}{\partial t} = E_{in} - E_\pm - i\theta E_\pm -i\eta\frac{\partial^2E_\pm}{\partial\tau^2} + i\left(A|E_\pm|^2E_\pm + B|E_\mp|^2E_\pm + 2A\langle|E_\pm|^2 \rangle E_\pm + B\langle |E_\mp|^2\rangle E_\pm + B\langle E_\pm E_\mp^*\rangle E_\mp\right),
\end{equation}
where the terms with angled brackets correspond to $\langle f \rangle = (1/t_R)\int_{-t_R/2}^{\;t_R/2} f\left(\tau\right) \,d\tau$, i.e. the temporal averages over the round-trip time $t_R$. The presence of this kind of averaged terms in Fabry-Pérot configurations was first suggested by Firth in \cite{firth81} when describing the phase shift due to the counter-propagating fields in the Kerr medium.

\section*{Data Availability}
The data that support the plots within this paper and other findings of this study are available from the corresponding author upon reasonable request.

\section*{Code Availability}
The codes that support the plots within this paper and other findings of this study are available from the corresponding author upon reasonable request.

\end{widetext}

\bibliography{References}

\section*{Acknowledgments}
LH acknowledges funding provided by the CNQO group within the Department of Physics at the University of Strathclyde, the ``Saltire Emerging Researcher" scheme through SUPA (Scottish University's Physics Alliance) and provided by the Scottish Government and Scottish Funding Council (SFC), and the SALTO funding scheme from the Max-Planck-Gesellschaft. This work was further supported by the European Union’s H2020 ERC Starting Grants 756966, the Marie Curie Innovative Training Network “Microcombs” 812818 and the Max-Planck-Gesellschaft.

\section*{Author Contributions}
LH and GLO defined the research project. LH completed the derivation of the used model with assistance from EMH \& GC. The numerical simulations were completed by LH \& EMH with assistance from GC. The stability analysis was completed by GC \& LH. All authors assisted with the analysis and discussions on the results. LH wrote the manuscript with inputs from all authors. PDH, GLO \& LH secured funding for the project.

\section*{Competing Interests}
The authors declare no competing interests.

\end{document}